\documentclass[twocolumn,twoside,slac]{revtex4}
\usepackage{graphicx}
\usepackage{fancyhdr}

\input{tcilatex}

\begin{document}

\title{How to Claim a Discovery}
\author{W.A. Rolke}
\affiliation{University of Puerto Rico - Mayaguez}
\author{A.M. L\'{o}pez}
\affiliation{University of Puerto Rico - Mayaguez}

\begin{abstract}
We describe a statistical hypothesis test for the presence of a signal. The
test allows the researcher to fix the signal location and/or width a priori,
or perform a search to find the signal region that maximizes the signal. The
background rate and/or distribution can be known or might be estimated from
the data. Cuts can be used to bring out the signal.
\end{abstract}

\maketitle

\thispagestyle{fancy}


\section{Introduction}

Setting limits for new particles or decay modes has been an active research
area for many years. In high energy physics it received renewed interest
with the unified method by Feldman and Cousins \cite{Cousins-Feldman}.
Giunti \cite{Giunti} and Roe and Woodroofe \cite{Roe-Woodroofe} gave
variations of the unified method, trying to resolve an apparent anomaly when
there are fewer events in the signal region than expected. They all discuss
the problem of setting limits for the case of a known background rate. The
case of an unknown background rate was discussed in a conference talk by
Feldman \cite{Feldman} and a method for handling this case was developed by
Rolke and L\'{o}pez \cite{Rolke and Lopez}. Little work has been done though
on the question of claiming a discovery. This problem could be handled by
finding a confidence interval and claiming a discovery if the lower limit is
positive. Instead the question of a discovery should be done separately, by
performing a hypothesis test with the null hypothesis $H_{o}$%
:\textquotedblright There is no signal present\textquotedblright . Rejecting
this hypothesis will then lead to a claim for a new discovery. In carrying
out a hypothesis test one needs to decide on the type I error probability $%
\alpha $, the probability of falsely rejecting the null hypothesis. This is
of course equivalent to the major mistake to be guarded against, namely that
of falsely claiming a discovery.

In practice a hypothesis test is often carried out by finding the p-value.
This is the probability that an identical experiment will yield a result as
extreme (with respect to the null hypothesis) or even more so given that the
null hypothesis is true. Then if $p<\alpha $\ we reject $H_{0}$; otherwise
we fail to do so. For the test discussed here it is not possible to compute
the p-value analytically, and therefore we will find the p-value via Monte
Carlo.

Maybe the most important decision in carrying out a hypothesis test is the
choice of $\alpha $, or what we might call the discovery threshold. As we
shall see, this decision is made much easier by the method described here
because we will need only one threshold, regardless of how the analysis was
done. What a proper discovery threshold should be in high energy physics is
a question outside the scope of this paper, although we might suggest $%
\alpha =0.001$\ (roughly equivalent to $3\sigma $). Sinervo \cite{Sinervo}
argues for a much stricter standard of $5\sigma $, or $\alpha =2.9\ast
10^{-7}$. We believe that such extreme values were used in the past because
it was felt that the calculated p values were biased downward by the
analysis process, and a small $\alpha $ was needed in order to compensate
for any unwittingly introduced biases. If we were to trust that our p-value
is in fact correct, a $1$\ in $1000$\ error rate should to be acceptable.

A general introduction to hypothesis testing with applications to high
energy physics is given in Sinervo \cite{Sinervo}. \ A classic reference for
the theory of hypothesis testing is Lehmann \cite{Lehmann}.

\section{The Signal Test}

Our test uses $T=x-b$\ or $T=x-y/\tau $\ as the test statistic, depending on
whether the background rate $b$\ is assumed to be known or not. Here $x$\ is
the number of observations in the signal region, $y$\ is the number of
observations in the background region and $\tau $\ is the probability that a
background event falls into the background region divided by the probability
that it falls into the signal region. Therefore $y/\tau $\ is the estimated
background in the signal region and $x-y/\tau $\ is an estimate for the
signal rate $\lambda $. $T$\ is the maximum likelihood estimator of $\lambda 
$, and it is the quantity used in Feldman and Cousins \cite{Cousins-Feldman}
without being set to $0$\ when $x-y/\tau $\ is negative. This is not
necessary here because a negative value of $x-y/\tau $\ will clearly lead to
a failure to reject $H_{0}$.

Other choices for the test statistic are of course possible. For example, a
measure for the size of a signal that is often used in high energy physics
is $S/\sqrt{b}$. Under the null hypothesis this statistic is approximately
Gaussian, at least if there is sufficient data. Unfortunately the
approximation is not sufficiently good in the extreme tails where a new
discovery is made, leading to p-values that are much smaller than is
warranted. Even when using Monte Carlo to compute the true p-value, this
test statistic can be shown to be inferior to the one proposed in our method
because it has consistently lower power, that is its probability of
detecting a real signal is smaller.

In order to find the p-value of the test we need to know the null
distribution. In the simplest case of a known background rate and everything
else fixed this is given by the Poisson distribution, but in all other cases
it is not possible to compute the null distribution analytically, and we
will therefore find it via Monte Carlo. As an illustration consider the
following case shown in figure 1: here we have $100$\ events on the interval 
$[0,1]$, with the signal region a priori set to be $[0.44,0.56]$. There are $%
25$\ events in the signal region, and the background distribution is known
to be flat. Therefore we find $x=25$, $y=75$, $\tau =7.33$\ and $T=14.77$.
Because we know that the background is flat on $[0,1]$, and because under
the null hypothesis all $100$\ events are background we can simulate this
experiment by drawing $100$ observations from a uniform distribution on $%
[0,1]$\ and computing $T$\ for this Monte Carlo data set. Repeating this $%
150000$\ times we find the histogram of Monte Carlo $T$\ values shown in
figure 2, case 1. In this simulation $8$\ of the $150000$\ simulation runs
had a value of $T$\ greater than $14.77$, or $p=0.000053$. Using $\alpha
=0.0001$\ we would therefore reject the null hypothesis and claim a
discovery. Note that in addition to rejecting the null hypothesis we can
also turn the p-value into a significance by using the Gaussian distribution
and claim that this signal is a $3.87\sigma $\ effect.

How would things change if the signal region had not been fixed a priori but
instead was found by searching through all signal regions centered at $0.5$\
and we would have accepted any signal with a width between $0.01$\ and $0.2$%
? That is if we had kept the signal location fixed but find the signal width
that maximizes $T$, the estimate of the number of signal events? Again we
can find the null distribution via Monte Carlo, repeating the exact analysis
for each simulation run individually. The histogram of $T$\ values for this
case is shown in figure 2, case 2. Here we find a value of $T$\ larger than $%
14.77$\ in $570$\ of the $150000$\ runs for a p-value of $0.0038$\ or $%
2.67\sigma $. At a discovery threshold of $\alpha =0.001$\ we would
therefore not find this signal significant anymore.

Even more, what if we also let the signal location vary, say anywhere in $%
[0.2,0.8]$? That is for any pair of values $(L,H)$\ we define $[L,H]$\ as
the signal region and $[0,L),(H,1]$\ as the background region, compute $%
T_{L,H}$\ for this pair and then maximize over all possible values of $L$\
and $H$. Note that because $T_{L,H}$\ is monotonically increasing in $\tau $
as long as all the observations stay either in the signal or in the
background region, we can find the maximum fairly quickly by letting $L$\
and $H$\ be the actual observations. The histogram of $T_{L,H}$\ values for
this case is shown in figure 2, case 3. We find a value of $T$\ larger than $%
14.77$\ in $9750$\ of the $150000$\ runs for a p-value of $0.065$\ or $%
1.51\sigma $, clearly not significant.

It was necessary in the second and third cases above to limit the search
region somewhat, to the interval $[0.2,0.8]$ and to signals at least $0.01$
and at most $0.2$ wide, because otherwise the largest value of $T$ is almost
always found for a very wide signal region, even when a clear narrow signal
is present. This restriction will not induce a bias as long as the decision
on where to search are made a priori.

In the general situation where the background is not flat on $[0,1]$\ we can
make use of the probability integral transform. Of course this requires
knowledge of the background distribution $F$, but if it is not known we can
estimate it from the data, either using a parametric function fitted to the
data or even using a nonparametric density estimator. Again all calculations
are done under the null hypothesis so we do not need to worry about the
signal or its distribution.

As long as we copy exactly for the Monte Carlo events what was done for the
real data we will find the correct p-value. This includes using cuts used to
improve the signal to noise ratio, but it then requires the ability to
correctly Monte Carlo all the variables used for cutting, including their
correlations.

\section{Performance of the Method}

As an illustration for the performance of the signal test consider the
following experiment: we generate $100$\ events from a linear background on $%
[3,5]$\ and (if present) a Gaussian signal at $3.9$\ with a width of $0.05$.
Then we find the signal through a variety of situations, from the one
extreme where everything is fixed a priori to the other where the largest
signal of any width is found. The background density is found by fitting and
the background rate is estimated. The\ power curves are shown in figure 3.
No matter what combination of items were fixed a priori or were used to
maximize the test statistic, and with it the signal to noise ratio, all
cases achieved the desired type I error probability, $\alpha =0.05$. Not
surprisingly the more items are fixed a priori, the better the power of the
test.

\section{Conclusion}

We have described a statistical hypothesis test for the presence of a
signal. Our test is conceptually simple and very flexible, allowing the
researcher a wide variety of choices during the analysis stage. It will
yield the correct type I error probability as long as the Monte Carlo used
to find the null distribution exactly mirrors the steps taken for the data.
Monte Carlo studies have shown that this method has satisfactory power.

\section{\protect\bigskip Acknowledgments}

\label{Ack}This work was partially supported by the Division of High Energy
Physics of the US Department of Energy.

\section{\protect\bigskip Appendix}

\FRAME{fbFU}{352.625pt}{275pt}{0pt}{\Qcb{100 Events on [0,1], with the
signal region a priori set to be [0.44, 0.56]. There are 25 events in the
signal region, and the background distribution is assumed to flat.}}{}{%
discoveryfig1.eps}{\special{language "Scientific Word";type
"GRAPHIC";maintain-aspect-ratio TRUE;display "USEDEF";valid_file "F";width
352.625pt;height 275pt;depth 0pt;original-width 7.7236in;original-height
5.9525in;cropleft "0";croptop "1.0186";cropright "1.0073";cropbottom
"0";filename 'rolkephystatfig1.EPS';file-properties "XNPEU";}}

\FRAME{ftbpFU}{352.625pt}{275pt}{0pt}{\Qcb{Histograms of T values of Monte
Carlo simulation.}}{}{discoveryfig2.eps}{\special{language "Scientific
Word";type "GRAPHIC";maintain-aspect-ratio TRUE;display "USEDEF";valid_file
"F";width 352.625pt;height 275pt;depth 0pt;original-width
7.7236in;original-height 5.9525in;cropleft "0";croptop "1.0161";cropright
"1.0047";cropbottom "0";filename 'rolkephystatfig2.EPS';file-properties
"XNPEU";}}

\FRAME{ftbpFU}{352.625pt}{275pt}{0pt}{\Qcb{Power curves for 10 different
cases such as signal location fixed a priori or not, same for signal width,
background estimated or ect. alpha=0.05 is used.}}{}{discoveryfig4.eps}{%
\special{language "Scientific Word";type "GRAPHIC";maintain-aspect-ratio
TRUE;display "USEDEF";valid_file "F";width 352.625pt;height 275pt;depth
0pt;original-width 7.7236in;original-height 5.9525in;cropleft "0";croptop
"0.9842";cropright "0.9731";cropbottom "0";filename
'rolkephystatfig3.EPS';file-properties "XNPEU";}}

\end{document}